\def\b{\beta}\def\d{\delta}
\def\h{\theta}
\def\o{\omega}\def
\p{\pi}\def\r{\rho}

\def\D{\Delta}

\def\inf{\infty}\def\ha{{1\over 2}}

\def\({\left(}\def\){\right)}\def\[{\left[}\def\]{\right]}

\def\coo{coordinates }

\def\rep{representation }

\def\ssy{spherically symmetric }

\def\pb{Poisson brackets }

\def\QM{quantum mechanics }

\def\cor{commutation relations }

\def\sys{symplectic structure }\def\sf{symplectic form }

\def\section#1{\bigskip\noindent{\bf#1}\smallskip}

\def\PL#1{Phys.\ Lett.\ {\bf#1}}

\def\PR#1{Phys.\ Rev.\ {\bf#1}}\def\CQG#1{Class.\ Quantum Grav.\ {\bf#1}}
\def\NP#1{Nucl.\ Phys.\ {\bf#1}}\def\GRG#1{Gen.\ Relativ.\ Grav.\ {\bf#1}}

\def\JoP#1{J.\ Phys.\ {\bf#1}} \def\IJMP#1{Int.\ J. Mod.\ Phys.\ {\bf #1}}

\def\EJP#1{Eur.\ J.\ Phys.\ {\bf#1}}

\def\arx#1{{\tt arXiv:#1}}

\def\ref#1{\medskip\everypar={\hangindent 2\parindent}#1}
\def\beginref{\begingroup
\bigskip
\centerline{\bf References}
\nobreak\noindent}
\def\endref{\par\endgroup}

\magnification=1200

{\nopagenumbers
\line{}
\vskip60pt
\centerline{\bf Spectrum of the hydrogen atom in Snyder space}
\smallskip
\centerline{\bf in a semiclassical approximation}

\vskip60pt
\centerline{
{\bf B. Iveti\'c}$^{1,2,}$\footnote{$^\dagger$}{e-mail: boris.ivetic@irb.hr}
{\bf S. Mignemi}$^{1,3,}$\footnote{$^\ddagger$}{e-mail: smignemi@unica.it},
and {\bf A. Samsarov}$^{1,3,}$\footnote{$^*$}{e-mail: samsarov@unica.it},}
\vskip10pt
\centerline{$^1$Dipartimento di Matematica e Informatica, Universit\`a di Cagliari}
\centerline{viale Merello 92, 09123 Cagliari, Italy}
\smallskip
\centerline{$^2$Rudjer Bo\v skovi\'c Institute, Bijeni\v cka c. 544, 10002 Zagreb, Croatia}
\smallskip
\centerline{$^3$INFN, Sezione di Cagliari, Cittadella Universitaria, 09042 Monserrato, Italy}
\vskip80pt
\centerline{\bf Abstract}
\medskip
{\noindent We study the spectrum of the hydrogen atom in Snyder space in a semiclassical
approximation based on a generalization of the Born-Sommerfeld quantization rule. While
the corrections to the standard quantum mechanical spectrum arise at first order in the
Snyder parameter for the $l=0$ states, they are of second order for $l\ne0$.
This can be understood as due to the different topology of the regions of integration in
phase space.}
\vskip10pt
{\noindent

}
\vskip80pt\
\vfil\eject}

\section{1. Introduction}
Quantum mechanics (QM) with modified \cor has attracted much attention
[1,2], since it implies the existence of a minimal observable length, in
accordance with most models of quantum gravity, that predict a minimal length of the
order of the Planck scale [3].

A particularly interesting example of deformation of the canonical \cor is given by
the nonrelativistic Snyder model [4,5],
which is based on the Euclidean three-dimensional version of the \cor originally
proposed by Snyder [6], in his search for a divergenceless field theory. These \cor
read
$$[x_i,p_j]=i(\d_{ij}+\b^2p_ip_j),\qquad[x_i,x_j]=i\b^2J_{ij},\qquad[p_i,p_j]=0,
\eqno(1)$$
where $J_{ij}=x_ip_j-x_jp_i$ are the generators of rotations and $\b$ is a
parameter with the dimension of inverse momentum.
The model admits an $SO(4)$ symmetry in phase space, generated by the generators
$J_{ij}$ and $x_i$.
Its implications have been investigated by several authors both in its classical
[7-8] and quantum version [4-5] and some
simple systems like the harmonic oscillator have been exactly solved.

The one-dimensional version of the Snyder \QM coincides with one of the favourite
models of deformed QM, its only nontrivial commutator being
$$[x,p]=i(1+\b^2p^2).\eqno(2)$$
This has been the subject of even wider investigations [2,4,9-12], showing in particular
that it implies  the existence  of a minimal resolution attainable from measures of
length, $\D x\ge\b$.

In this letter, we deal with the problem of the hydrogen atom in Snyder space.
This problem has been studied in several papers [10-14].
In one dimension, it is possible to obtain exact results for the
energy spectrum [10-12], while in the more interesting three-dimensional case, only
perturbative solutions have been found [13-14]. Moreover, while in one dimension the
corrections to the spectrum are of order $\b$ [10-12]\footnote{$^1$} {The result
of [10], where the first corrections are found at order $\b^2$, is not correct, see
[11].}, in three dimensions they start from order $\b^2$ [13-14], and hence the zero
angular momentum sector of
the theory, which coincides with the 1D problem, cannot be obtained as a smooth
limit of the higher angular momentum sector for $l\to0$.
This fact seems to have been disregarded in previous investigations\footnote{$^2$}
{Actually, perturbative calculations give divergent results for $l=0$ [14]. This appears
as a confirmation that in this case the perturbative expansion does not start from order
$\b^2$ terms.}.

The reason of this discrepancy lays presumably in the fact that the three-dimensional
Hamiltonian
usually postulated for the Snyder model does not preserve its full $SO(4)$ symmetry,
but only its rotational subgroup, and hence the degeneracy in the angular momentum
of the QM hydrogen spectrum is lost in this case. The full symmetry could be
restored by modifying the potential in a suitable way [15,8].

In this paper, we obtain an exact solution for the spectrum of the hydrogen
atom in Snyder space in a semiclassical approximation, using the Bohr-Sommerfeld
(BS) quantization rule [16]. In this way the problem is essentially reduced to a
classical one. We remark that in the 1D case and also in
3D ordinary QM this calculation gives the exact spectrum. We do not know whether
this is true also for 3D Snyder QM, but certainly one obtains at least a good
approximation.

To perform the calculation, polar momentum coordinates must be used. These are not
common in classical
mechanics, but are the equivalent of those used in the momentum \rep of QM,
and are simply the dual of the usual polar position coordinates.
The use of a momentum \rep is standard in the investigation of the Snyder dynamics,
because it gives the most natural \rep of the Snyder \cor [3,4].

In standard QM, the investigation of the hydrogen atom in momentum representation
has a long history. It was first investigated by Fock [17]. The one-dimensional
case had later been considered in [18].

\section{2. The semiclassical approximation}
The semiclassical approximation starts from the study of the classical dynamics,
where the commutators are replaced by Poisson brackets.
Since the \sys of the Snyder model is noncanonical, its classical dynamics is
best written in Hamiltonian form. The standard choice for the Hamiltonian is
identical to the classical one,
$$H={{\bf p}^2\over2m}-{e^2\over r},\eqno(3)$$
where ${\bf p}^2$ is the square of the momentum and $r=\sqrt{{\bf x}^2}$.
This Hamiltonian is invariant under the rotation group $SO(3)$, but not under
the full $SO(4)$ Snyder group [15]. However, it is the one that is usually adopted
in this context. Due to the rotational invariance, the angular momentum is conserved
and hence the classical orbits are confined to a plane, as in Newtonian mechanics,
and the quantum spectrum will be independent on the magnetic quantum number.
In future, we plan to study more general Hamiltonians that preserve the full Snyder
group.

The difference between Snyder and Newtonian mechanics resides in the symplectic
structure, that is of course modified. For example, in cartesian coordinates, the
Snyder \sf can be written as
$$d\o_S=-\(\d_{ij}-{\b^2p_ip_j\over1+\b^2p^2}\)x_idp_j.\eqno(4)$$
This leads to a different classical motion in the two cases.
\medskip
To calculate the spectrum of the 3D hydrogen atom in a semiclassical approximation,
we have to generalize the standard BS quantization condition to noncanonical systems.
We require that the phase integral over each degree of freedom is an integer,
$$\oint d\o_i=2\p n_i,\eqno(5)$$
where $d\o=\sum_id\o_i$ is the \sf relative to the model under study.
In the Newtonian case (5) reduces to the standard rule since $d\o_{N,i}=p_idx_i$.

Before discussing the 3D problem, we briefly review the results obtained in 1D in
[11], using a slightly different approach from ours.
Of course, in 1D the rotational invariance is absent and the problem simplifies.

The 1D Hamiltonian is
$$H={p^2\over2m}-{e^2\over x}=-E,\eqno(6)$$
with $x>0$, where the constant $-E$ is the total energy (the minus sign is inserted
because we are considering bound states, so that $E$ is positive).
The Newtonian \sf is simply $d\o_N=pdx$ and the BS condition reads
$$\oint pdx=-\oint xdp=2\p n.\eqno(7)$$
In the following, we shall use the second form, because it is more apt to a
generalization to the Snyder case.

On a closed orbit, starting from the origin, $p$ goes from $\inf$ to 0 and in the way back
from $0$ to $-\inf$, while (6) yields $x=2me^2/(p^2+2mE)$.
One has then
$$\oint d\o_N=-\oint xdp=\int^\inf_{-\inf}{2me^2\over{p^2+2mE}}\ dp=\p\sqrt{2me^4\over E}.
\eqno(8)$$
Imposing the BS condition gives the spectrum
$$E_n={me^4\over2n^2},\eqno(9)$$
which coincides with the exact result [18].

In the Snyder case, the calculation is exactly the same, but the \sf is
$$d\o_S=-{xdp\over1+\b^2p^2}.\eqno(10)$$
The generalized BS quantization condition reads then
$$\int^\inf_{-\inf}{2me^2dp\over(p^2+2mE)(1+\b^2p^2)}={\sqrt{2me^4}\ \p\over\sqrt E(1+\b\sqrt{2mE})}
=2\p n.\eqno(11)$$
This again coincides with the exact condition found in [11].
Note that a "miracolous" cancellation turns the dependence  of the corrections on $\b^2$,
that one would expect from the integral in (11), into a dependence on $\b$.
More explicitly, an expansion in powers of $\b$ gives
$$E_n={me^4\over2n^2}\(1-{2\b me^2\over n}+\dots\).\eqno(12)$$

\bigbreak
\section{3. The hydrogen atom in three dimensions}
In three dimensions, the BS rule requires that the integral along an orbit of each term of the \sf
must be an integer. The calculation is best performed using spherical coordinates in the space
of momenta. The BS condition applied to the angular variables gives as usual the quantization of the
$z$-component $\tt m$ and of the norm $l$ of the angular momentum.

For what concerns the radial part, one can simplify the calculation, recalling that because of the
conservation of the angular momentum, valid both in the Newtonian and Snyder cases, the classical
motion is confined to a plane. We shall therefore use polar momentum coordinates on a plane,
defined as [19]
$$\eqalign{&
p_\r=\sqrt{p_1^2+p_2^2},\qquad p_\h=\arctan{p_2\over p_1},\cr
&\r={p_1x_1+p_2x_2\over\sqrt{p_1^2+p_2^2}},\qquad J=J_{12}=x_1p_2-x_2p_1.}$$
These are duals of the polar position coordinates. In particular, $p_\r$ is the
norm of the momentum $\bf p$ and $\r$ is not the radial coordinate, but rather its
projection along $\bf p$. $J$ is the angular momentum, that is conserved for \ssy potentials
and takes the value $l$.

The previous \coo obey the \pb
$$\eqalign{&\{p_\r,p_\h\}=0,\qquad\quad\ \{\r,J\}=0,\qquad\quad\ \{p_\r,J\}=0,\cr
&\{p_\r,\r\}=1+\b^2p_\r^2,\qquad\{p_\h,J\}=1,\qquad\{p_\h,\r\}=0,}$$
from which one can easily obtain the \sys $d\o$, which maintains a simple form when passing
from Newtonian to Snyder mechanics [19]. In fact,
$$d\o_N= -(\r dp_\r+Jdp_\h)\to d\o_S=-\({\r dp_\r\over1+\b^2p_\r^2}+Jdp_\h\).\eqno(14)$$
However, in these \coo the Hamiltonian takes an unusual form,
$$H={p_\r^2\over2m}-{e^2\over\sqrt{\r^2+J^2/p_\r^2}}=-E\eqno(15)$$
where $E$ is the conserved energy.

For $l=0$, all the formulae reduce to the ones valid in 1D, and the spectrum is given by (12).
If $l\ne0$, instead,
$$\r=\pm l\ {\sqrt{-p_\r^4+({4m^2e^4\over l^2}-4mE)p_\r^2-4m^2E^2}\over p_\r(p_\r+2mE)},\eqno(16)$$
and $\r$ is real in the interval $z_-\le z\le z_+$, where we have defined $z=p_\r^2$, and
$$z_\pm=2m\({me^4\over l^2}-E\pm{e^2\over l}\sqrt{{m^2e^4\over l^2}-2mE}\).\eqno(17)$$

In the Newtonian case, the radial integral reads
$$\oint-\r dp_\r=l\int_{z_-}^{z_+}{\sqrt{(z-z_-)(z_+-z)}\over z(z+2mE)}\ dz=\p\(\sqrt{2me^4\over E}-2l\).
\eqno(18)$$
Notice that, contrary to the 1D case, the integral extends over a finite region of phase space.
Equating the result to $2\p n$ and defining a new quantum number $n'=n+l$, one obtains
the well-known spectrum $E_{n'}={me^4\over2n'^2}$.

In the Snyder case, the only change is in the symplectic form. We have now
$$\eqalign{&\oint-{\r dp_\r\over1+\b^2p_\r^2}=l\int_{z_1}^{z_2}{\sqrt{(z-z_1)(z_2-z)}\over z(z+2mE)
(1+\b^2z)}\ dz\cr
&=\p\[{\sqrt{2me^4}\over\sqrt{E}(1-2\b^2mE)}-l\(1+\sqrt{1+{4\b^2 m^2e^4\over l^2(1-2\b^2mE)}}\)\].}
\eqno(19)$$
Equating (19) to $2\p n$, one finds that the corrections to the Newtonian result are now of order $\b^2$,
namely
$$E_{n'l}={me^4\over2n'^2}\[1+{2\b^2m^2e^4\over n'}\({1\over n'}-{1\over l}\)+\dots\].\eqno(20)$$

Note that, as expected, due to the breaking of the Snyder symmetry, the spectrum
is no longer degenerate in the angular quantum number $l$.
The result is in good agreement with the perturbative calculation performed in [14], that
predicts
$$E_{n'l}={me^4\over2n'^2}\[1+{2\b^2m^2e^4\over n'}\({1\over n'}-{1\over l+\ha}+{1\over2l(l+\ha)(l+1)}\)
+\dots\].$$
We remark that the perturbative result breaks down for $l=0$. This is a signal that in that case the
first order corrections are not of order $\b^2$.

The different behaviour of the spectrum for nonvanishing angular momentum with respect to $l=0$
seems to arise because the region of integration in the phase integrals is topologically different in
the two cases and cannot be smoothly deformed.
The corrections have opposite sign and different order of magnitude in the two cases; we are not
aware of similar behaviour in other physical systems.

It would be interesting to study the symmetry group of the Snyder hydrogen atom, and to
investigate if suitably modifying the potential in order to recover the
symmetry under the full Snyder group, also the standard hydrogen atom symmetry is recovered, so that
the difference between $l=0$ and $l\ne0$ disappears.
This topic is presently under study.
\bigbreak
\section{Acknowledgements}
This work was partially supported by the European Commission and the Croatian Ministry of Science,
Education and Sports through grant project financed under the Marie Curie FP7-PEOPLE-2011-COFUND,
project NEWFELPRO.

\beginref
\ref [1] G. Veneziano, Europhys. Lett. {\bf2}, 199 (1986);
M. Maggiore, \PL{B304}, 63 (1993); \PL{B319}, 83 (1993).
\ref [2] A. Kempf, G. Mangano and R. Mann, \PR{D52}, 1108 (1995).
\ref [3] For a review, see L.J. Garay, \IJMP{A10}, 145 (1993).
\ref [4] S. Mignemi, \PR{D84}, 025021 (2011).
\ref [5] Lei Lu and A. Stern, \NP{B854}, 894 (2011); \NP{B860}, 186 (2012).
\ref [6] H.S. Snyder, \PR{71}, 38 (1947).
\ref [7] L.N. Chang, D. Mini\'c, N. Okamura and T. Takeuchi, \PR{D65}, 125027 (2002);
S. Benczik, L.N. Chang, D. Mini\'c, N. Okamura, S. Rayyan and T. Takeuchi, \PR{D66}, 026003 (2002).
\ref [8] B. Iveti\'c, S. Meljanac and S. Mignemi, \CQG{31}, 105010 (2014).
\ref [9] K. Nozari and T. Azizi, \GRG{38} 38, 735 (2006); P. Pedram, \IJMP{19}, 2003 (2010).
\ref [10] R. Akhoury and Y.P. Yao, \PL{B572}, 37 (2003).
\ref [11] T.V. Fityo, I.O. Vakarchuk and V.M. Tkachuk, \JoP{A39}, 2143 {2006}
\ref [12] D. Bouaziz and N. Ferkous, \PR{A82}, 022105 (2010).
\ref [13] F. Brau, \JoP{A32}, 7691 (1999).
\ref [14] S. Benczik, L.N. Chang, D. Mini\'c and T. Takeuchi, \PR{A72}, 012104 (2005).
\ref [15] S. Mignemi, \CQG{29}, 215019 (2012).
\ref [16] A. Sommerfeld, Sitz. Bayer. Akad. Wiss. M\"unchen 425 (1915).
\ref [17] V.A. Fock, Zs. Phys. {\bf98}, 145 (1935).
\ref [18] H.N. N\'u\~{n}ez Y\'epez, C.A. Vargas and A.L. Salas Brito, \EJP{8}, 189 (1987).
\ref [19] S. Mignemi and R. \v Strajn, \arx{1501.01447}.
\endref
\end